\documentclass[aps,prl,twocolumn,reprint,superscriptaddress,nobibnotes]{revtex4-2}
\usepackage{amssymb, amsmath}
\usepackage{adjustbox}
\usepackage{graphicx}
\usepackage{dcolumn}
\usepackage{multirow}
\usepackage{bm}
\usepackage[T1]{fontenc}
\usepackage{setspace}
\usepackage{physics}
\usepackage{natbib}
\usepackage{mathrsfs}
\usepackage{float}
\usepackage{siunitx}
\usepackage{hyperref}
\usepackage{booktabs}
\usepackage{tabularx}
\hypersetup{colorlinks=true, citecolor=blue, urlcolor=blue, linkcolor=blue, filecolor=gray}
\usepackage{tikz,xcolor}
\definecolor{lime}{HTML}{A6CE39}
\DeclareRobustCommand{\orcidicon}{%
	\begin{tikzpicture}
		\draw[lime, fill=lime] (0,0) 
		circle [radius=0.16] 
		node[white] {{\fontfamily{qag}\selectfont \tiny ID}}; \draw[white, fill=white] (-0.0625,0.095) 
		circle [radius=0.007];    \end{tikzpicture}
	\hspace{-2mm}}
\foreach \x in {A, ..., Z}{%
	\expandafter\xdef\csname orcid\x\endcsname{\noexpand\href{https://orcid.org/\csname orcidauthor\x\endcsname}{\noexpand\orcidicon}}
}

\begin{document}

\title{Multiple Layer-Selective Polar Charge Density Waves in ${\rm{EuTe}}_{4}$}

\author{Wen-Han Dong\orcidA{}}
\affiliation {State Key Laboratory of Low Dimensional Quantum Physics and Department of Physics, Tsinghua University, Beijing  100084, China}

\author{Wenhui Duan\orcidB{}}
\affiliation {State Key Laboratory of Low Dimensional Quantum Physics and Department of Physics, Tsinghua University, Beijing 100084, China}
\affiliation {Chen-Ning Yang Institute for Advanced Study, Tsinghua University, Beijing 100084, China}
\affiliation {Frontier Science Center for Quantum Information, Beijing, China}
\affiliation {Suzhou National Laboratory, Suzhou 215123, China}

\author{Yong Xu\orcidC{}}
\email {yongxu@mail.tsinghua.edu.cn}
\affiliation {State Key Laboratory of Low Dimensional Quantum Physics and Department of Physics, Tsinghua University, Beijing 100084, China}
\affiliation {Frontier Science Center for Quantum Information, Beijing, China}
\affiliation {Suzhou National Laboratory, Suzhou 215123, China}

\author{Peizhe Tang\orcidD{}}
\email {peizhet@buaa.edu.cn}
\affiliation {School of Materials Science and Engineering, Beihang University, Beijing 100191, China}
\affiliation {Max Planck Institute for the Structure and Dynamics of Matter, Center for Free-Electron Laser Science, 22761 Hamburg, Germany}


\begin{abstract}
${\rm{EuTe}}_{4}$ is a polar charge density wave (CDW) material, with giant thermal hysteresis and non-volatile state switching under electric and optical fields, attracting great attention in recent years. However, the in-depth understanding of these anomalous phenomena remains elusive. Herein, via first-principles calculations, we reveal that the polar CDW state in ${\rm{EuTe}}_{4}$ hosts a novel layer-selective nature, wherein multiple energetically close CDW configurations coexist and exhibit low interconversion energy barriers. Monte Carlo simulations indicate that the giant thermal hysteresis in ${\rm{EuTe}}_{4}$ originates from a configurational-entropy-driven crossover between ordered and disordered states, around which the material hosts a metastable CDW state characterized by diverse local polar configurations breaking the out-of-plane translational symmetry. The configurational composition of this metastable CDW state can be effectively controlled by electric and optical fields, thereby enabling non-volatile state switching. Our theoretical findings align well with recent experimental observations in ${\rm{EuTe}}_{4}$ and pave the way for exploring the emerging phenomena and applications of polar CDW in multilayered systems.
\end{abstract}

\maketitle

\emph{Introduction---}The coexistence and interplay of charge density wave (CDW) \cite{CDW1,CDW2,TMD1} with other ordered states such as superconductivity, magnetism, and pair density wave have attracted significant interest in the field of condensed matter physics \cite{Fradkin15,TMD2,Kagome1,Kagome2,Kagome3,Kagome4,Kagome5}. If the point-group or time-reversal symmetries are broken, the CDW states may intertwine with polar or chiral orders, yielding polar or chiral CDWs \cite{Polar1,Polar2,Polar3,Polar4,Chiral1,Chiral2,Chiral3}. Among these, polar CDW is a unique platform for exploring the fundamental physics at the intersection of CDW and ferroelectrics. By combining optical tunability \cite{CDW3,CDW4,CDW5,CDW6} and ferroelectric controllability \cite{FO1,FO2}, polar CDW opens avenues for next-generation optoelectronics and non-volatile memory.

\begin{figure}[ht!]
	\centering
	\includegraphics[width=8.6cm]{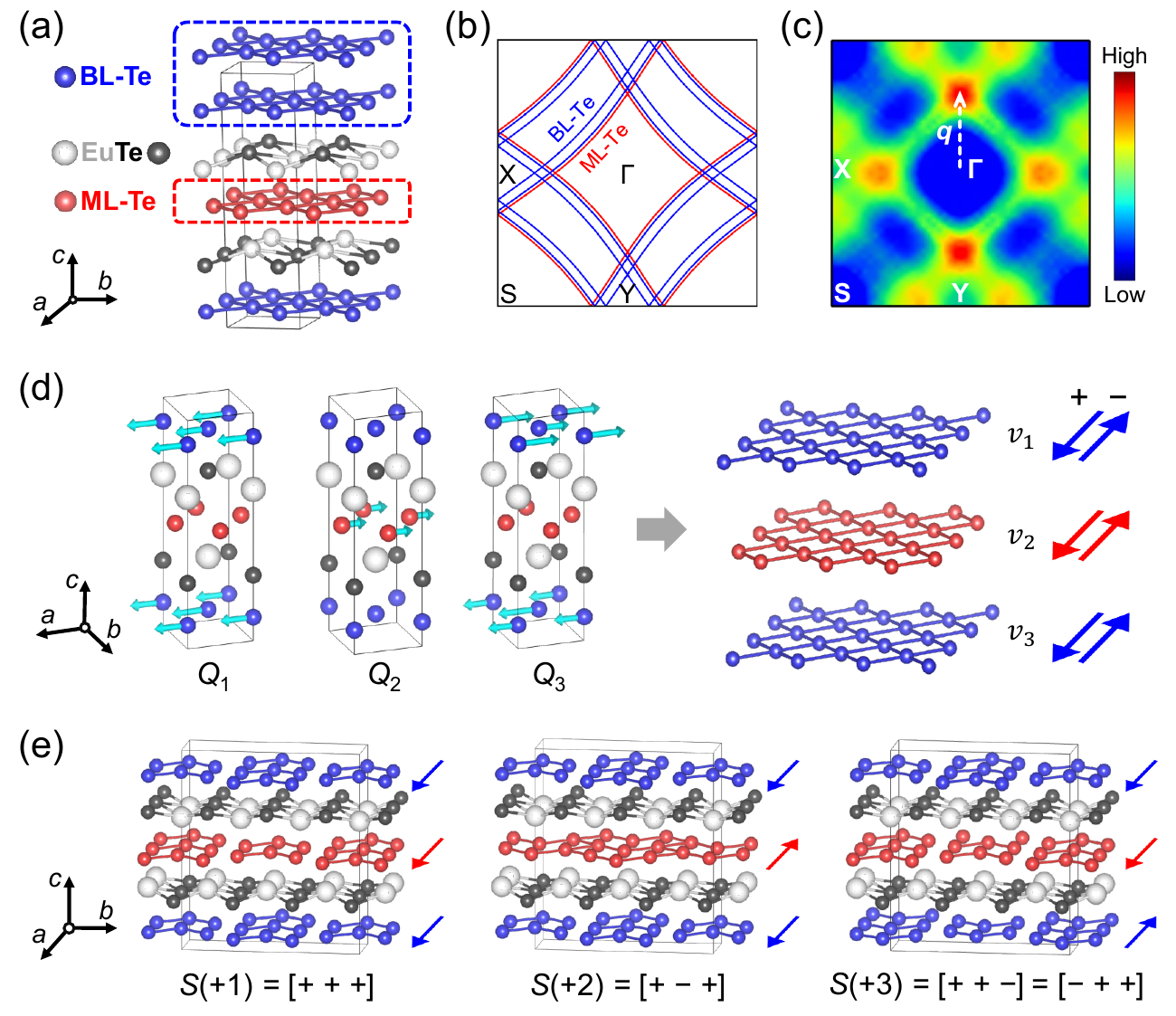}
	\caption{Multiple layer-selective polar CDW configurations in ${\rm{EuTe}}_{4}$. (a) Crystal structure of non-CDW ${\rm{EuTe}}_{4}$ in $Pmmn$ symmetry, showing relatively weak couplings between Te layers. (b) Fermi surface of non-CDW ${\rm{EuTe}}_{4}$ from tight-binding model. (c) Lindhard function corresponding to (b), exhibiting a dominant peak around $\boldsymbol{q} = \frac{1}{3}\boldsymbol{b^*}$. (d) Schematics for the imaginary phonons $Q_{1,2,3}$ at $\boldsymbol{q} = \frac{1}{3}\boldsymbol{b^*}$ (left) and the three-layer CDW representations $[\nu_{1}\;\nu_{2}\;\nu_{3}]$ (right), where $\nu_{1,2,3}=\pm$ describes the polar order of each Te layer. (e) Three nonequivalent $1\times3\times1$ CDW configurations with overall positive polarization: $S(+1)$, $S(+2)$, and $S(+3)$. Their counterparts with negative polarization are $S(-1) = [---]$, $S(-2) = [-+-]$, and $S(-3) = [--+]=[+--]$, respectively.}
	\label{fig1}
\end{figure}

Recently, the multilayered material ${\rm{EuTe}}_{4}$ has garnered great attention due to the coexistence of CDW and polar orders \cite{Yang23,Wu19,LV22,LV24,LV25,Electrical,Optical,Meng22,ThermoE25,Oh25,Xiao24,Bansal23,CALC22,Ning25,Nonlocal23}. 
Its structural unit consists of a Te monolayer (ML-Te) and a Te bilayer (BL-Te), separated by EuTe layers [see Fig.~\ref{fig1}(a)], and exhibits structural anisotropy with distinct in-plane lattice constants \cite{Wu19}. This material hosts a near-commensurate unidirectional CDW within its Te layers \cite{Wu19,LV22,LV24,LV25}, which persists up to the transition temperature $T_{\mathrm{CDW}}\sim 652$ K \cite{Bansal23}. The CDW coexists with polar order, as confirmed by second-harmonic generation detection \cite{Optical}. Interestingly, below $T_{\mathrm{CDW}}$, ${\rm{EuTe}}_{4}$ displays a giant thermal hysteresis in resistance, with a hysteresis loop spanning over 400 K \cite{Wu19,LV22,Yang23}. Within the thermal hysteresis loop, the CDW state with polar order can be electrically \cite{Electrical} and optically \cite{Optical} switched in a non-volatile manner at room temperature, suggesting promising applications of ${\rm{EuTe}}_{4}$ as efficient memristor \cite{Electrical} and laser-controlled memory \cite{Optical}. Prior theoretical studies suggest that the observed CDW in ${\rm{EuTe}}_{4}$ at low temperature is driven by the electron-phonon coupling \cite{Wu19,Xiao24,Bansal23,CALC22}. While, as the temperature increases, the origins of multiple metastable CDW states and rich emergent phenomena---such as giant thermal hysteresis and non-volatile state switching under electric and optical fields---are yet to be well understood. Therefore, the theoretical understanding of the temperature evolution of CDW states in ${\rm{EuTe}}_{4}$ is urgently needed, which is also crucial for advancing future research and applications of polar CDW.

In this Letter, through density functional theory (DFT) calculations, we reveal that ${\rm{EuTe}}_{4}$ hosts a novel layer-selective polar CDW as the ground state, where charge density modulations are selectively driven by the $a$-axial polar atomic displacements within both ML-Te and BL-Te. At finite temperature, multiple layer-selective polar CDW configurations coexist in ${\rm{EuTe}}_{4}$, which are energetically close due to the relatively weak interlayer couplings between these Te layers and exhibit low interconversion energy barriers. Furthermore, we develop an effective model based on DFT results and perform Monte Carlo (MC) simulations to study the thermal evolution of these CDW configurations. Our simulation results reveal that a metastable CDW state comprising multiple local polar configurations emerges with a configurational-entropy-driven crossover between ordered and disordered states, thereby leading to a giant thermal hysteresis upon heating and cooling. Electric and optical fields can alter the configurational compositions in such a metastable CDW state, varying its configurational entropy and thus enabling non-volatile state switching. Our work proposes a new mechanism to fully understand recent experimental discoveries in ${\rm{EuTe}}_{4}$ and opens new avenues for exploring polar CDW in multilayered systems.

\emph{Layer-selective polar CDW---}We begin by elucidating the origin of layer-selective CDW in ${\rm{EuTe}}_{4}$, where "layer-selective" means CDW occurs selectively within Te layers. Prior DFT studies reveal that the quasi-2D bands in ${\rm{EuTe}}_{4}$ around the Fermi level are mainly contributed by ML-Te and BL-Te \cite{Xiao24,CALC22}. To identify the layer contributions to CDW states in ${\rm{EuTe}}_{4}$, we construct a tight-binding model considering the intralayer/interlayer interaction inside ML-Te and BL-Te while ignoring the interlayer coupling between them, which well reproduces the DFT electronic structures (see Fig.~S1 in Supplemental Material (SM) \cite{SI}). Figure~\ref{fig1}(b) shows the Fermi surface calculated from the tight-binding model, with overlapping Fermi pockets from ML-Te and BL-Te due to their close onsite energies (Fig.~S2 in SM \cite{SI}). Influenced by in-plane anisotropy, these Fermi pockets together induce a pronounced unidirectional CDW instability around $\boldsymbol{q} = \frac{1}{3}\boldsymbol{b^*}$ in the Lindhard function [see Fig.~\ref{fig1}(c)], matching the experimentally observed $\boldsymbol{q}_{\mathrm{CDW}}\sim 0.33\boldsymbol{b^*}$ \cite{Wu19} and suggesting electron-phonon coupling as the origin of CDW \cite{Nesting}. Layer-resolved Lindhard functions further indicate that the CDW forms within each Te layer (see Fig.~S2 in SM \cite{SI}), indicating a layer-selective CDW behavior that accounts for the coexisting CDWs discovered in ML-Te and BL-Te \cite {LV24,LV25}. We highlight the uniqueness of this unidirectional, layer-selective CDW in ${\rm{EuTe}}_{4}$ in contrast to other rare-earth tellurides, see Sec.~III of SM \cite{SI}.

\begin{table}[b]
	\begin{center}
		\caption{Nonequivalent $1\times3\times2$ CDW configurations, corresponding to the paired configurations for neighboring sites in MC simulations. Here, the linkage "$+$" means out-of-plane stacking of $1\times3\times1$ structural units. The energy is defined per $1\times3\times1$ supercell, which contains 6 Eu and 24 Te atoms.}
		\label{tbl:table1}
		\fontsize{8.0pt}{8.0pt}
		\vspace{0.3cm}
		\begin{tabular*}{\linewidth}{@{\extracolsep{\fill}}lccr@{}} 
			\toprule
			\toprule
			Configuration & $S(+1)$+$S(+1)$ & $S(+2)$+$S(+2)$ & $S(+3)$+$S(+3)$  \\		
			\midrule
			Energy (meV) & 30.03 & 37.80 & 6.72\\
			\midrule
			Configuration & $S(+1)$+$S(-1)$ & $S(+2)$+$S(-2)$ & $S(+3)$+$S(-3)$  \\
			\midrule
			Energy (meV) & \textbf{0.00} & 12.36 & 37.44\\
			\midrule
			Configuration & $S(+1)$+$S(+2)$ & $S(+1)$+$S(+3)$ & $S(+2)$+$S(+3)$ \\
			\midrule
			Energy (meV) & 33.00 & 16.41 & 21.90\\
			\midrule
			Configuration & $S(+1)$+$S(-2)$ & $S(+1)$+$S(-3)$ & $S(+2)$+$S(-3)$  \\
			\midrule
			Energy (meV) & 5.58 & 17.34 & 22.17\\
			\bottomrule	
			\bottomrule	
		\end{tabular*}
	\end{center}
\end{table}

To understand how polarization emerges from CDW-induced lattice distortion in ${\rm{EuTe}}_{4}$, we perform DFT calculations to study its phonon spectrum. Consistent with prior calculations \cite{Xiao24,CALC22}, we observe the imaginary phonons with lowest frequencies near $\boldsymbol{q} = \frac{1}{3}\boldsymbol{b^*}$ (see Fig.~S3 in SM \cite {SI}), supporting the Lindhard function results discussed above. Figure~\ref{fig1}(d) shows the vibration modes of three imaginary phonons $Q_{1,2,3}$ at $\boldsymbol{q} = \frac{1}{3}\boldsymbol{b^*}$. These phonons exhibit $a$-axial ($x$-directional) polar atomic displacements within ML-Te or BL-Te, with $B_{2}$ ($x$-polarized) or $A_{2}$ (anti-$x$-polarized) irreducible representations of the $C_{2v}$ little group at $\boldsymbol{q} = \frac{1}{3}\boldsymbol{b^*}$. Driven by these phonon instabilities, the CDW distortions in ${\rm{EuTe}}_{4}$ involve specific combinations of polar atomic displacements within each Te layer. Consequently, the CDW intrinsically coexists with an $a$-axial polar order in each Te layer [see Fig.~\ref{fig1}(d)], indicating that the CDW and polarization have a common microscopic lattice origin and underscoring a layer-selective polar CDW feature.

Due to the relatively weak interlayer couplings between Te layers, multiple energetically close, layer-selective polar CDW configurations should coexist in ${\rm{EuTe}}_{4}$. Guided by possible combinations of imaginary phonons $Q_{1,2,3}$, we employ DFT calculations to identify possible commensurate CDW configurations. As shown in Fig.~\ref{fig1}(e), we discover three distinct ferroelectric CDW configurations and their counterparts within the minimal $1\times3\times1$ supercell. For simplicity, these unit configurations are abbreviated as $S(\pm1)$, $S(\pm2)$ and $S(\pm3)$, with $+$ ($-$) denoting an overall polarization along $x$ ($-x$) direction. The calculated polarizations for these configurations are 2.20, 2.62 and 3.56 ($\times10^{-3}$ C/$\rm{m}^{2}$), respectively \cite{SI}. Once we further consider the couplings between structural units and enlarge the supercell along $z$ direction (i.e., $1\times3\times2$ supercell), all possible low-energy CDW configurations are listed in Table~\ref{tbl:table1}, which are energetically close (maximal energy difference 1.26 meV/atom), with the antiferroelectric $S(+1)$+$S(-1)$ being the ground-state configuration. Notably, these CDW configurations exhibit low interconversion energy barriers ($<$ 5.4 meV/atom), indicating that structural transitions can occur among them as temperature increases and multiple local configurations could coexist in the CDW state below $T_{\mathrm{CDW}}$ (see Fig.~S5 in SM \cite{SI}).

\begin{figure}[htp!]
	\centering
	\includegraphics[width=8.6cm]{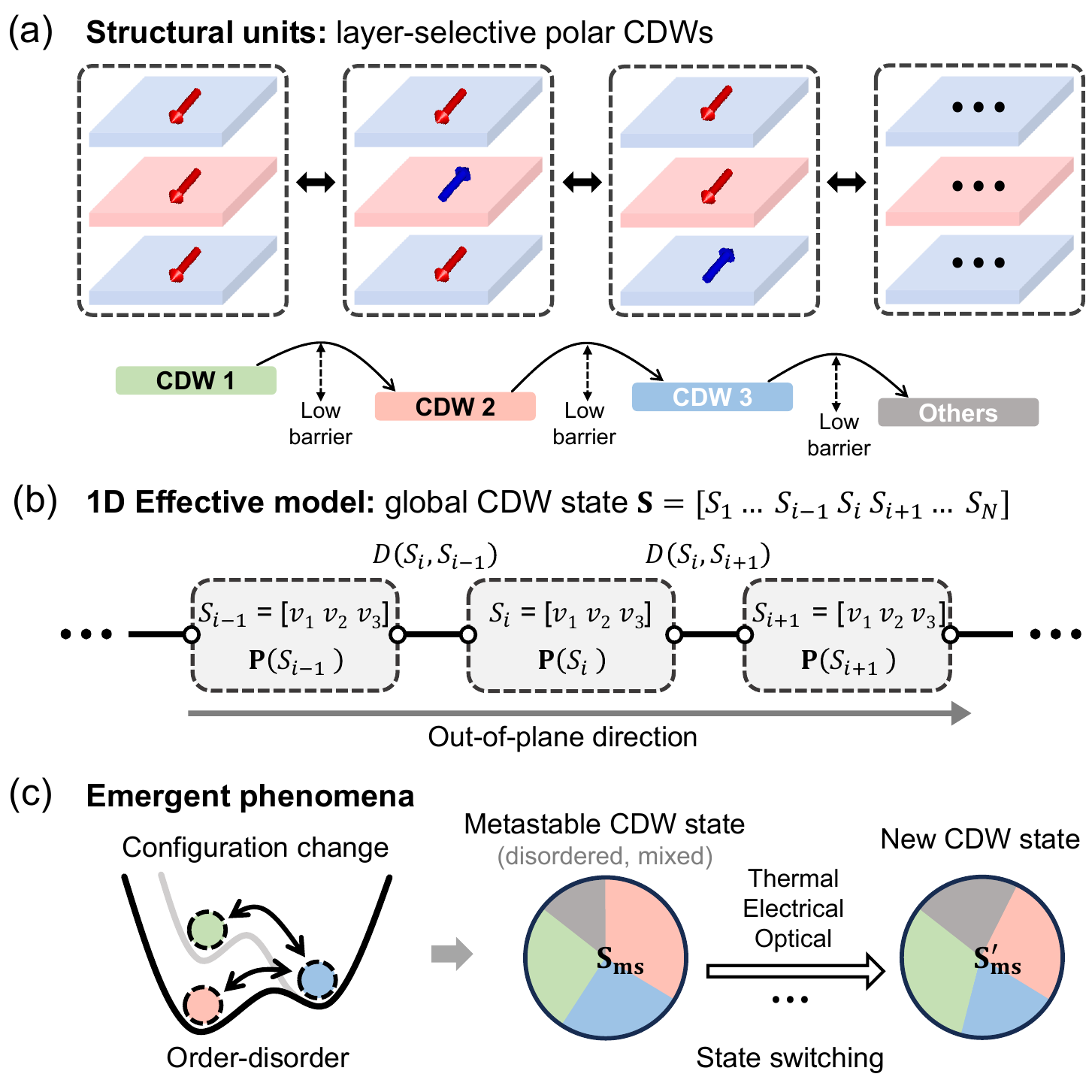}
	\caption{Framework for understanding ${\rm{EuTe}}_{4}$-like systems. (a) Multiple layer-selective polar CDW configurations with low interconversion energy barriers. (b) Schematic for the effective model, where each site hosts one $1\times3\times1$ CDW configuration among $S(\pm1)$, $S(\pm2)$, and $S(\pm3)$. (c) Schematics for the underlying order-disorder physics (left) and non-volatile state switching (right). The balls depict various possible paired configurations, while fan areas show their populations in the metastable CDW state.}
	\label{fig2}
\end{figure}

\emph{Effective model for MC simulations---}To simulate the thermal effects associated with multiple energetically close CDW configurations in ${\rm{EuTe}}_{4}$, we develop a 1D effective model based on the  DFT-informed low-energy physics, see elaboration in Sec.~V of SM \cite{SI}. In this model, each unit site corresponds to a $1\times3\times1$ supercell that adopts one possible CDW configuration among $S(\pm1)$, $S(\pm2)$, and $S(\pm3)$ [see Fig.~\ref{fig2}(a)], with probabilities determined by their relative energies and temperature. We assume the polarization of each CDW configuration is temperature-independent, while thermal fluctuations only affect the system's entropy, further influencing the statistical distribution of local configurations.

Along the out-of-plane direction, the whole system has $N$ sites, described by a global CDW state $\mathbf{S}=[S_{1}\;S_{2}\;...\;S_{N-1}\;S_{N}]$. Each site $i$ hosts a local configuration $S_{i}$ with polarization $\mathbf{P}(S_{i})$, and couples to adjacent sites via nearest-neighbor interactions $D(S_{i},S_{i\pm 1})$ [see Fig.~\ref{fig2}(b)]. The averaged Landau-Devonshire-type internal energy $U(\mathbf{S})$ \cite{FE1,GL2001} is:

\begin{small}
	\begin{equation}
		\label{eq1}
		\begin{split}
			&U(\mathbf{S})=\frac{1}{N}\sum^{N}_{i}\{\frac{A(S_{i})}{2}\mathbf{P}^{2}(S_{i})
			+\frac{B(S_{i})}{4}\mathbf{P}^{4}(S_{i})
			\\&+\frac{C(S_{i})}{6}\mathbf{P}^{6}(S_{i})+\frac{D(S_{i},S_{i\pm 1})}{2}(\mathbf{P}(S_{i})-\mathbf{P}(S_{i\pm 1}))^{2}\},
		\end{split}
	\end{equation}
\end{small}

{\setlength{\parindent}{0pt}\setlength{\parskip}{0.75\baselineskip}where the first three terms describe the polarization-contributed energy within each site, and the last term captures inter-site interactions between neighboring sites. The local polarization $\mathbf{P}(S_{i})$ and parameters $A(S_{i})$, $B(S_{i})$, $C(S_{i})$ and $D(S_{i},S_{i\pm 1})$ are fitted from DFT calculations (see Tables~S1 and S2 in SM \cite{SI}). Since the global CDW state could host multiple local configurations, we introduce the averaged Helmholtz free energy $H(\mathbf{S})$ to include entropy contributions \cite{Entropy}:}

\begin{small}
	\begin{equation}
		\label{eq2}
		H(\mathbf{S})=U(\mathbf{S})-TS_{\mathrm{config}},\,S_{\mathrm{config}}=-k_{\mathrm{B}}\sum_{\mathbb{C}}{n_{\mathbb{C}}}\ln(\frac{n_{\mathbb{C}}}{N_{\mathbb{C}}}).
	\end{equation}
\end{small}

{\setlength{\parindent}{0pt}\setlength{\parskip}{0.75\baselineskip}Here, $T$ is temperature. $S_{\mathrm{config}}$ is configurational entropy, while the effect of vibrational entropy is estimated to be negligible, see Sec.~VI of SM \cite{SI}. $k_{\mathrm{B}}$ is Boltzmann constant. $n_{\mathbb{C}}$ is the population of each possible paired configuration for two neighboring sites [see Table~\ref{tbl:table1}]. In a 1D model with $N$ sites, there are $N$ paired configurations in total. $N_{\mathbb{C}}$ is the degeneracy for each paired configuration, where $N_{\mathbb{C}}=4$ for the pairs with $S(+3)$+$S(\pm3)$, $N_{\mathbb{C}}=2$ for that with $S(+1)$+$S(\pm3)$ and $S(+2)$+$S(\pm3)$, and $N_{\mathbb{C}}=1$ for others. Note that this definition of $S_{\mathrm{config}}$ inherently includes contributions from all possible stacking configurations.}

\begin{figure}[htb]
	\centering
	\includegraphics[width=8.6cm]{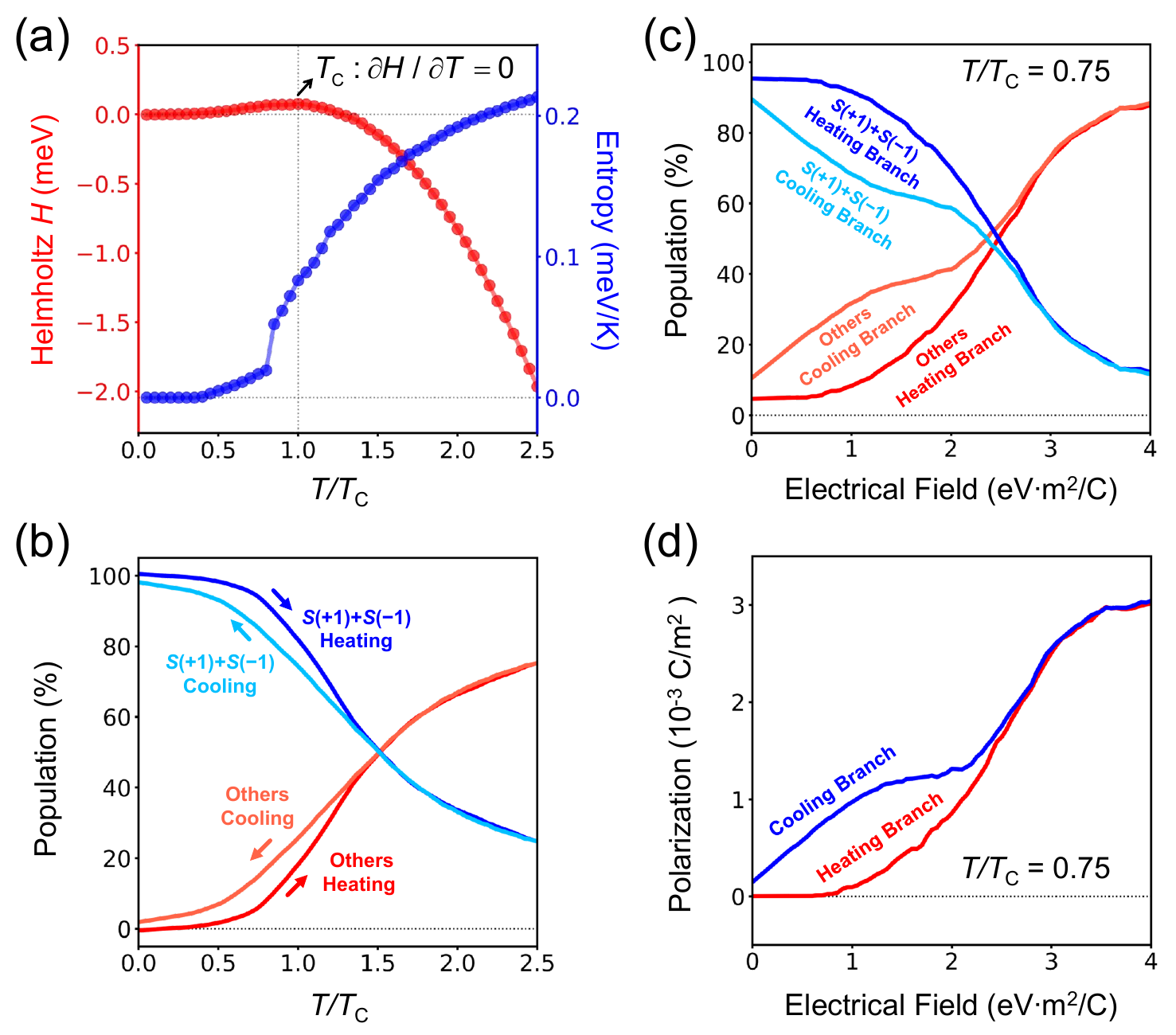}
	\caption{Thermal hysteresis and electrical state control. (a) Temperature-dependent Helmholtz free energy $H(\mathbf{S})$ (red) and configurational entropy $S_{\mathrm{config}}$ (blue), obtained from \textit{average-mode} MC simulations \cite{SI}. Here, the temperature ($T$) is normalized by crossover temperature ($T_{\mathrm{C}}$). (b) Population ratios of the ground-state paired configuration ($S(+1)$+$S(-1)$) versus other configurations ("Others") upon heating and (then) cooling, obtained from \textit{heating-cooling-mode} MC simulations \cite{SI} with a sweep rate of $|\delta(T/T_{\mathrm{C}})| = 5.0\times 10^{-3}$. (c) Population ratios of paired configurations and (d) polarization under an $x$-directional electric field, obtained from \textit{average-mode} MC simulations \cite{SI} and initialized from the heating and cooling branches of (b) at $T/T_{\mathrm{C}} = 0.75$. In (b-d), the curves are smoothed to reduce fluctuations.}
	\label{fig3}
\end{figure}

Having established the effective model, we proceed to explore thermal effects in ${\rm{EuTe}}_{4}$ using MC simulations, see computational details in SM \cite{SI}. At zero temperature, the system adopts an ordered, antiferroelectric ground state $\mathbf{S}_{\mathbf{g}}$, in which any neighboring sites exhibit a paired configuration as $S(+1)$+$S(-1)$. As temperature increases, thermal excitation can transform the local paired configurations in $\mathbf{S}_{\mathbf{g}}$ into other ones with close energies [see Fig.~\ref{fig2}(c)], inducing a metastable CDW state $\mathbf{S}_{\mathbf{ms}}$ that breaks translation symmetry along the out-of-plane direction. Hereafter, our MC simulations will substantiate the thermally driven $\mathbf{S}_{\mathbf{ms}}$ in ${\rm{EuTe}}_{4}$ and illustrate that the properties of $\mathbf{S}_{\mathbf{ms}}$ are controllable thermally, electrically, and optically, yielding rich emergent phenomena as depicted in Fig.~\ref{fig2}(c).

\emph{MC simulation results---}We first simulate the thermal behavior of ${\rm{EuTe}}_{4}$ to uncover the underlying physical origin for the giant thermal hysteresis observed experimentally \cite{Wu19,LV22,Yang23}. Figure~\ref{fig3}(a) presents the temperature dependence of $H(\mathbf{S})$ and $S_{\mathrm{config}}$. Below the crossover temperature  $T_{\mathrm{C}}$, $H(\mathbf{S})$ and $S_{\mathrm{config}}$ slightly increase with temperature as the system accumulates CDW fluctuations deviating from the ground state to gradually accommodate compositional disorder (see Fig.~S7 in SM \cite{SI}). Upon approaching $T_{\mathrm{C}}$, $H(\mathbf{S})$ decreases gradually and $S_{\mathrm{config}}$ increases sharply, indicating a crossover between ordered and disordered states in ${\rm{EuTe}}_{4}$ \cite{Entropy}. This crossover reflects the melting of the ordered $\mathbf{S}_{\mathbf{g}}$ into $\mathbf{S}_{\mathbf{ms}}$, which contains diverse local configurations with layer-selective polar CDW at high temperature. To reveal the thermal hysteresis resulting from this order-disorder crossover, we simulate the thermal evolution of the global CDW state in a heating-cooling cycle initiated from $\mathbf{S}_{\mathbf{g}}$, while tracking the change of populations for the ground-state paired configuration and other configurations. Our results show a notable thermal hysteresis loop centered around $T_{\mathrm{C}}$ [see Fig.~\ref{fig3}(b) and Fig.~S8 in SM \cite{SI}]. To distinguish $\mathbf{S}_{\mathbf{ms}}$ in different thermal branches, we denote them as $\mathbf{S}^{\mathbf{h}}_{\mathbf{ms}}$ and $\mathbf{S}^{\mathbf{c}}_{\mathbf{ms}}$ for the heating and cooling branches, respectively. Over a broad temperature range, $\mathbf{S}^{\mathbf{h}}_{\mathbf{ms}}$ and $\mathbf{S}^{\mathbf{c}}_{\mathbf{ms}}$ exhibit markedly different CDW configuration distributions and $S_{\mathrm{config}}$, yielding a pronounced resistivity difference between the heating and cooling branches, consistent with the giant thermal hysteresis observed in transport measurements (see resistivity fitting in Fig.~S13~\cite{SI}). Thus, our findings promise to resolve the long-standing debate about the origin of the giant thermal hysteresis in ${\rm{EuTe}}_{4}$ (see justification in Sec.~VIII of SM \cite{SI}) and provide a new perspective on the thermal hysteretic phenomena in CDW materials beyond known mechanisms, such as impurity pinning \cite{Impurity}, sliding stacking order \cite{TaS219}, metal-to-insulator transition \cite {MTI}, and CDW transitions across wave vectors \cite{MultiQ1,MultiQ2}.

\begin{figure}[b!]
	\centering
	\includegraphics[width=8.6cm]{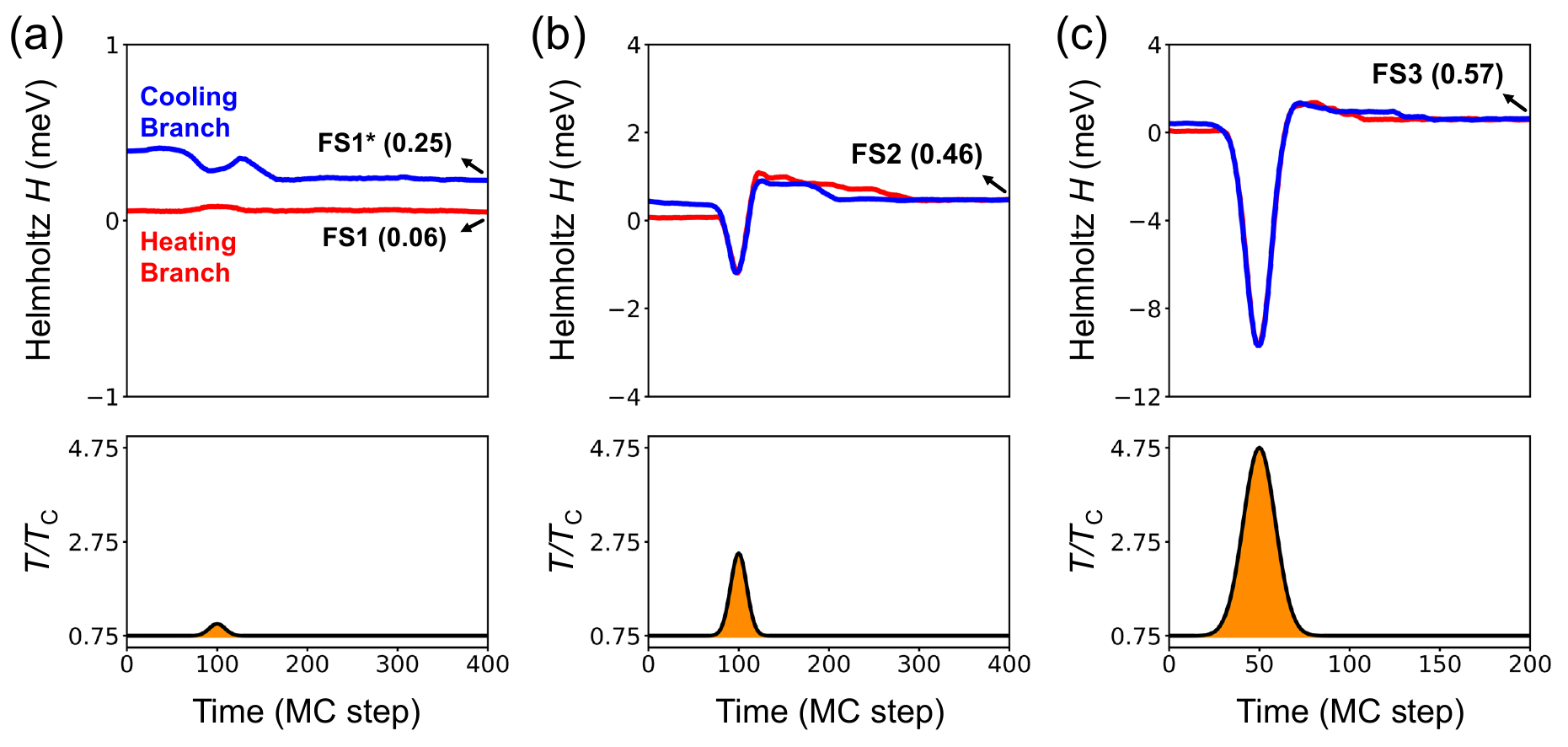}
	\caption{Optical state control. (a) MC-simulated evolution of $H(\mathbf{S})$ (upper) under a Gaussian heat pulse with amplitude of 0.25 (bottom). (b) and (c) display the results under pulse amplitudes of 1.75 and 4, respectively. In (a-c), the curves are smoothed and are initialized from the heating and cooling branches of Fig.~\ref{fig3}(b) at $T/T_{\mathrm{C}} = 0.75$. FS$\ast$ denote the final states, with their $H(\mathbf{S})$ shown in parentheses.}
	\label{fig4}
\end{figure}

Within the thermal hysteresis loop, the CDW state in ${\rm{EuTe}}_{4}$ can be manipulated electrically \cite{Electrical}. To study the switching of $\mathbf{S}_{\mathbf{ms}}$ under a static electric field $\mathbf{E}$, we incorporate the electric-field-polarization coupling term $-\frac{1}{N}\sum^{N}_{i}\mathbf{E}\cdot \mathbf{P}(S_{i})$ \cite{GL2001} into $H(\mathbf{S})$ in Eq.~\ref{eq2}. Figure~\ref{fig3}(c) shows an efficient electrical control of $\mathbf{S}_{\mathbf{ms}}$ at $T/T_{\mathrm{C}} = 0.75$, where a notable difference persists between $\mathbf{S}^{\mathbf{h}}_{\mathbf{ms}}$ and $\mathbf{S}^{\mathbf{c}}_{\mathbf{ms}}$ [see Fig.~\ref{fig3}(b)]. Under an increasing electric field, the switching of $\mathbf{S}_{\mathbf{ms}}$ occurs smoothly, indicating a continuous reorganization of the system's local polar configurations, as also reflected by the non-linear, gradual increase of polarization in Fig.~\ref{fig3}(d). This finding provides essential insights into the smooth, non-volatile electrical switching of resistance observed in ${\rm{EuTe}}_{4}$ \cite{Electrical}, confirming that its metastable CDW states consist of numerous local polar configurations, thereby producing electrically controllable free-energy landscapes. Furthermore, Fig.~\ref{fig3}(d) shows that, under a relatively low electric field ($< 2.2~\rm{eV \cdot m^2 / C}$), the cooling branch exhibits larger polarization than the heating branch, revealing the energy storage during thermal hysteresis in ${\rm{EuTe}}_{4}$ and suggesting promising electrocaloric applications \cite{EC25,EC23,EC21}.

Inspired by experimental observations of photoinduced long-lived metastable CDW states in ${\rm{EuTe}}_{4}$ \cite{Optical}, we investigate the optical control of $\mathbf{S}_{\mathbf{ms}}$. Since the observed features of equilibrium states are insensitive to light polarization and coherence \cite{Optical}, we conclude the optical pumping as a photoinduced thermal effect and simulate the optical fields via the Gaussian heat pulses. Figure~\ref{fig4} shows that, at $T/T_{\mathrm{C}} = 0.75$, the photoinduced modulations of $\mathbf{S}^{\mathbf{h}}_{\mathbf{ms}}$ and $\mathbf{S}^{\mathbf{c}}_{\mathbf{ms}}$ can be tuned by varying the pulse strength. When we apply a weak pumping pulse, final states [FS1 and FS1* in Fig.~\ref{fig4}(a)] remain close to their unpumped states. As we increase the pumping strength, a relatively strong pulse first thermally excites $\mathbf{S}^{\mathbf{h}}_{\mathbf{ms}}$ and $\mathbf{S}^{\mathbf{c}}_{\mathbf{ms}}$ into high-$S_{\mathrm{config}}$ states, after which they relax to similar final states [FS2 in Fig.~\ref{fig4}(b)]. Once we further enhance the pumping strength, new final states with higher $H(\mathbf{S})$ [FS3 in Fig.~\ref{fig4}(c)] can be achieved. We propose a physical picture for these optical processes in Fig.~S10 of SM \cite{SI}, suggesting that photoinduced thermal effects can significantly tailor the system's hysteresis behavior, enabling hidden metastable CDW states that are inaccessible via thermal excitation. Our simulations illuminate the underlying mechanism to explain the pulse-dependent, non-volatile optical state switching in ${\rm{EuTe}}_{4}$ \cite{Optical}, which would guide non-volatile memory applications in this and similar CDW materials.

\emph{Conclusion---}Our first-principles calculations identify ${\rm{EuTe}}_{4}$ as a layer-selective polar CDW material, in which multiple energetically close CDW configurations coexist around the ground state. Our MC simulations reveal that, as temperature increases, thermal excitation could alter the population of local paired CDW configurations and enhance the configurational entropy.  Hence, a crossover between ordered and disordered states can be achieved and result in a notable hysteretic behavior, which promisingly resolves the long-standing debate about the origin of giant thermal hysteresis in ${\rm{EuTe}}_{4}$. Within the thermal hysteresis loop, we further demonstrate that $\mathbf{S}_{\mathbf{ms}}$ can be effectively controlled under electric and optical fields, aligning with recent observations of non-volatile switching of CDW states in ${\rm{EuTe}}_{4}$ \cite{Electrical,Optical} and underscoring its promising applications in non-volatile memory \cite{Optical}, memristors \cite{Memristor}, thermoelectrics \cite{ThermoE25}, and electrocalorics \cite{EC25,EC23,EC21}. Future experimental validations of our proposed mechanism are outlined in Sec.~IX of SM \cite{SI}. This work not only clarifies the essential nature of CDW states in ${\rm{EuTe}}_{4}$, but also establishes a general framework for exploring a new class of polar CDW in multilayered systems.

\begin{acknowledgments}
	\emph{Acknowledgments---}We thank Dong Wu, Tao Dong, Qiaomei Liu and Jin Zhang for helpful discussions. This work was supported by the Basic Science Center Project of NSFC (Grant No. 52388201), the National Natural Science Foundation of China (Grants No. 12334003, No. 12421004, No. 12361141826, No. 12234011, No. 12374053 and No. 12504078), the National Key Basic Research and Development Program of China (Grants No. 2023YFA1406400 and No. 2024YFA1409100), the Fundamental and Interdisciplinary Disciplines Breakthrough Plan of the Ministry of Education of China (Grant No. JYB2025XDXM408), Quantum Science and Technology-National Science and Technology Major Project (Grants No. 2023ZD0300500 and No. 2021ZD0301800), Beijing Municipal Science and Technology Commission and Administrative Commission of Zhongguancun Science Park (Grant No. Z251100003625025), the China Postdoctoral Science Foundation (Grant No. 2026T190901), ShanghaiTech AI4S Initiative SHTAI4S202504, Beijing Key Laboratory of Quantum AI, the funding from Suzhou National Laboratory (SK-1902-2026-004), and Beijing Advanced Innovation Center for Future Chip (ICFC). The work was carried out at the National Supercomputer Center in Tianjin using the Tianhe new generation supercomputer.
\end{acknowledgments}
\nocite{*}

\bibliography{EuTe4}

@article{CDW1,
author = {Xuetao Zhu and Yanwei Cao  and Jiandi Zhang  and E. W. Plummer  and Jiandong Guo},
title = {Classification of charge density waves based on their nature},
journal = {Proc. Natl. Acad. Sci.},
volume = {112},
number = {8},
pages = {2367-2371},
year = {2015},
url = {https://www.pnas.org/doi/abs/10.1073/pnas.1424791112}
}

@book{CDW2,
  title={Charge density waves in solids},
  author={Gor'kov, Lev Petrovich and Gr{\"u}ner, George},
  volume={25},
  year={2012},
  publisher={Elsevier}
}

@article{TMD1,
url = {https://dx.doi.org/10.1088/0953-8984/23/21/213001},
year = {2011},
month = {may},
publisher = {},
volume = {23},
number = {21},
pages = {213001},
author = {Rossnagel, K},
title = {On the origin of charge-density waves in select layered transition-metal dichalcogenides},
journal = {J. Phys. Condens. Mat.}
}

@article{Fradkin15,
  title = {Colloquium: Theory of intertwined orders in high temperature superconductors},
  author = {Fradkin, Eduardo and Kivelson, Steven A. and Tranquada, John M.},
  journal = {Rev. Mod. Phys.},
  volume = {87},
  issue = {2},
  pages = {457--482},
  numpages = {26},
  year = {2015},
  month = {May},
  publisher = {American Physical Society},
  doi = {10.1103/RevModPhys.87.457},
  url = {https://link.aps.org/doi/10.1103/RevModPhys.87.457}
}

@article{TMD2,
  title={Discovery of a Cooper-pair density wave state in a transition-metal dichalcogenide},
  author={Liu, Xiaolong and Chong, Yi Xue and Sharma, Rahul and Davis, JC Seamus},
  journal={Science},
  volume={372},
  number={6549},
  pages={1447--1452},
  year={2021},
  url={https://www.science.org/doi/abs/10.1126/science.abd4607}
}

@article{Kagome1,
  title={AV3Sb5 kagome superconductors},
  author={Wilson, Stephen D and Ortiz, Brenden R},
  journal={Nat. Rev. Mater.},
  volume={9},
  number={6},
  pages={420--432},
  year={2024},
  url={https://www.nature.com/articles/s41578-024-00677-y}
}

@article{Kagome2,
    author = {Jiang, Kun and Wu, Tao and Yin, Jia-Xin and Wang, Zhenyu and Hasan, M Zahid and Wilson, Stephen D and Chen, Xianhui and Hu, Jiangping},
    title = {Kagome superconductors AV3Sb5 (A = K, Rb, Cs)},
    journal = {Nat. Sci. Rev.},
    volume = {10},
    number = {2},
    pages = {nwac199},
    year = {2022},
    month = {09},
    issn = {2095-5138},
    url = {https://doi.org/10.1093/nsr/nwac199}
}

@article{Kagome3,
  title={Discovery of charge density wave in a kagome lattice antiferromagnet},
  author={Teng, Xiaokun and Chen, Lebing and Ye, Feng and Rosenberg, Elliott and Liu, Zhaoyu and Yin, Jia-Xin and Jiang, Yu-Xiao and Oh, Ji Seop and Hasan, M Zahid and Neubauer, Kelly J and others},
  journal={Nature},
  volume={609},
  number={7927},
  pages={490--495},
  year={2022},
  url = {https://www.nature.com/articles/s41586-022-05034-z}
}

@article{Kagome4,
  title={Quantum states and intertwining phases in kagome materials},
  author={Wang, Yaojia and Wu, Heng and McCandless, Gregory T and Chan, Julia Y and Ali, Mazhar N},
  journal={Nat. Rev. Phys.},
  volume={5},
  number={11},
  pages={635--658},
  year={2023},
  url = {https://www.nature.com/articles/s42254-023-00635-7}
}

@article{Kagome5,
  title={Roton pair density wave in a strong-coupling kagome superconductor},
  author={Chen, Hui and Yang, Haitao and Hu, Bin and Zhao, Zhen and Yuan, Jie and Xing, Yuqing and Qian, Guojian and Huang, Zihao and Li, Geng and Ye, Yuhan and others},
  journal={Nature},
  volume={599},
  number={7884},
  pages={222--228},
  year={2021},
  url={https://www.nature.com/articles/s41586-021-03983-5}
}

@article{Polar1,
  title = {Charge-Stripe Order in the Electronic Ferroelectric ${\mathrm{LuFe}}_{2}{\mathrm{O}}_{4}$},
  author = {Zhang, Y. and Yang, H. X. and Ma, C. and Tian, H. F. and Li, J. Q.},
  journal = {Phys. Rev. Lett.},
  volume = {98},
  issue = {24},
  pages = {247602},
  numpages = {4},
  year = {2007},
  month = {Jun},
  publisher = {American Physical Society},
  url = {https://link.aps.org/doi/10.1103/PhysRevLett.98.247602}
}

@article{Polar2,
  title = {Polar charge and orbital order in $2H$-TaS${}_{2}$},
  author = {van Wezel, Jasper},
  journal = {Phys. Rev. B},
  volume = {85},
  issue = {3},
  pages = {035131},
  numpages = {5},
  year = {2012},
  month = {Jan},
  publisher = {American Physical Society},
  url = {https://link.aps.org/doi/10.1103/PhysRevB.85.035131}
}

@article{Polar3,
  title={Electrical switching of ferro-rotational order in nanometre-thick 1T-TaS2 crystals},
  author={Liu, Gan and Qiu, Tianyu and He, Kuanyu and Liu, Yizhou and Lin, Dongjing and Ma, Zhen and Huang, Zhentao and Tang, Wenna and Xu, Jie and Watanabe, Kenji and others},
  journal={Nat. Nanotechnol.},
  volume={18},
  number={8},
  pages={854--860},
  year={2023},
  url={https://www.nature.com/articles/s41565-023-01403-5}
}

@article{Polar4,
  title={Polar charge density wave in a superconductor with crystallographic chirality},
  author={Wu, Shangfei and Huang, Fei-Ting and Xu, Xianghan and Ritz, Ethan T and Birol, Turan and Cheong, Sang-Wook and Blumberg, Girsh},
  journal={Nat. Commun.},
  volume={15},
  number={1},
  pages={9276},
  year={2024},
  url={https://www.nature.com/articles/s41467-024-53627-1}
}

@article{Chiral1,
  title = {Chiral Charge-Density Waves},
  author = {Ishioka, J. and Liu, Y. H. and Shimatake, K. and Kurosawa, T. and Ichimura, K. and Toda, Y. and Oda, M. and Tanda, S.},
  journal = {Phys. Rev. Lett.},
  volume = {105},
  issue = {17},
  pages = {176401},
  numpages = {4},
  year = {2010},
  month = {Oct},
  publisher = {American Physical Society},
  url = {https://link.aps.org/doi/10.1103/PhysRevLett.105.176401}
}

@article{Chiral2,
  title={Chirality locking charge density waves in a chiral crystal},
  author={Li, Geng and Yang, Haitao and Jiang, Peijie and Wang, Cong and Cheng, Qiuzhen and Tian, Shangjie and Han, Guangyuan and Shen, Chengmin and Lin, Xiao and Lei, Hechang and others},
  journal={Nat. Commun.},
  volume={13},
  number={1},
  pages={2914},
  year={2022},
  url={https://www.nature.com/articles/s41467-022-30612-0}
}

@article{Chiral3,
  title={Atomic-scale visualization of chiral charge density wave superlattices and their reversible switching},
  author={Song, Xuan and Liu, Liwei and Chen, Yaoyao and Yang, Han and Huang, Zeping and Hou, Baofei and Hou, Yanhui and Han, Xu and Yang, Huixia and Zhang, Quanzhen and others},
  journal={Nat. Commun.},
  volume={13},
  number={1},
  pages={1843},
  year={2022},
  url={https://www.nature.com/articles/s41467-022-29548-2}
}

@article{CDW3,
  title = {The dynamics of charge-density waves},
  author = {Gr\"uner, G.},
  journal = {Rev. Mod. Phys.},
  volume = {60},
  issue = {4},
  pages = {1129--1181},
  numpages = {0},
  year = {1988},
  month = {Oct},
  publisher = {American Physical Society},
  url = {https://link.aps.org/doi/10.1103/RevModPhys.60.1129}
}

@article{CDW4,
  title={Fast electronic resistance switching involving hidden charge density wave states},
  author={Vaskivskyi, I and Mihailovic, IA and Brazovskii, S and Gospodaric, J and Mertelj, T and Svetin, D and Sutar, P and Mihailovic, D},
  journal={Nat. Commun.},
  volume={7},
  number={1},
  pages={11442},
  year={2016},
  url={https://www.nature.com/articles/ncomms11442}
}

@article{CDW5,
  title={Ultrafast manipulation of mirror domain walls in a charge density wave},
  author={Zong, Alfred and Shen, Xiaozhe and Kogar, Anshul and Ye, Linda and Marks, Carolyn and Chowdhury, Debanjan and Rohwer, Timm and Freelon, Byron and Weathersby, Stephen and Li, Renkai and others},
  journal={Sci. Adv.},
  volume={4},
  number={10},
  pages={eaau5501},
  year={2018},
  url={https://www.science.org/doi/full/10.1126/sciadv.aau5501}
}

@article{CDW6,
  title={Light-induced charge density wave in LaTe3},
  author={Kogar, Anshul and Zong, Alfred and Dolgirev, Pavel E and Shen, Xiaozhe and Straquadine, Joshua and Bie, Ya-Qing and Wang, Xirui and Rohwer, Timm and Tung, I-Cheng and Yang, Yafang and others},
  journal={Nat. Phys.},
  volume={16},
  number={2},
  pages={159--163},
  year={2020},
  url={https://www.nature.com/articles/s41567-019-0705-3}
}

@article{FO1,
  title={Towards two-dimensional van der Waals ferroelectrics},
  author={Wang, Chuanshou and You, Lu and Cobden, David and Wang, Junling},
  journal={Nat. Mater.},
  volume={22},
  number={5},
  pages={542--552},
  year={2023},
  url={https://www.nature.com/articles/s41563-022-01422-y#citeas}
}

@article{FO2,
  title={Ferroelectric order in van der Waals layered materials},
  author={Zhang, Dawei and Schoenherr, Peggy and Sharma, Pankaj and Seidel, Jan},
  journal={Nat. Rev. Mater.},
  volume={8},
  number={1},
  pages={25--40},
  year={2023},
  url={https://www.nature.com/articles/s41578-022-00484-3}
}

@article{Wu19,
  title = {Layered semiconductor ${\mathrm{EuTe}}_{4}$ with charge density wave order in square tellurium sheets},
  author = {Wu, D. and Liu, Q. M. and Chen, S. L. and Zhong, G. Y. and Su, J. and Shi, L. Y. and Tong, L. and Xu, G. and Gao, P. and Wang, N. L.},
  journal = {Phys. Rev. Mater.},
  volume = {3},
  issue = {2},
  pages = {024002},
  numpages = {7},
  year = {2019},
  month = {Feb},
  publisher = {American Physical Society},
  url = {https://link.aps.org/doi/10.1103/PhysRevMaterials.3.024002}
}

@article{Electrical,
  title={Room-temperature memristive switching between charge density wave states},
  author={Venturini, R and Rupnik, M and Ga{\v{s}}perlin, J and Lipi{\v{c}}, J and {\v{S}}utar, P and Vaskivskyi, Y and {\v{S}}{\v{c}}epanovi{\'c}, F and Grabnar, D and Gole{\v{z}}, D and Mihailovic, D},
  journal={Nat. Commun.},
  volume={17},
  number={1},
  pages={8360},
  year={2026},
  url={https://www.nature.com/articles/s41467-026-73267-x}
}

@article{Optical,
  title={Room-temperature non-volatile optical manipulation of polar order in a charge density wave},
  author={Liu, Qiaomei and Wu, Dong and Wu, Tianyi and Han, Shanshan and Peng, Yiran and Yuan, Zhihong and Cheng, Yihan and Li, Bohan and Hu, Tianchen and Yue, Li and others},
  journal={Nat. Commun.},
  volume={15},
  number={1},
  pages={8937},
  year={2024},
  url = {https://www.nature.com/articles/s41467-024-53323-0}
}

@article{LV22,
  title = {Unconventional Hysteretic Transition in a Charge Density Wave},
  author = {Lv, B. Q. and Zong, Alfred and Wu, D. and Rozhkov, A. V. and Fine, Boris V. and Chen, Su-Di and Hashimoto, Makoto and Lu, Dong-Hui and Li, M. and Huang, Y.-B. and others},
  journal = {Phys. Rev. Lett.},
  volume = {128},
  issue = {3},
  pages = {036401},
  numpages = {7},
  year = {2022},
  month = {Jan},
  publisher = {American Physical Society},
  url = {https://link.aps.org/doi/10.1103/PhysRevLett.128.036401}
}

@article{LV24,
  title = {Coexistence of Interacting Charge Density Waves in a Layered Semiconductor},
  author = {Lv, B. Q. and Zong, Alfred and Wu, Dong and Nie, Zhengwei and Su, Yifan and Choi, Dongsung and Ilyas, Batyr and Fichera, Bryan T. and Li, Jiarui and Baldini, Edoardo and others},
  journal = {Phys. Rev. Lett.},
  volume = {132},
  issue = {20},
  pages = {206401},
  numpages = {7},
  year = {2024},
  month = {May},
  publisher = {American Physical Society},
  url = {https://link.aps.org/doi/10.1103/PhysRevLett.132.206401}
}

@article{LV25,
  title={Large moir{\'e} superstructure of stacked incommensurate charge density waves},
  author={Lv, B. Q. and Su, Yifan and Zong, Alfred and Liu, Qiaomei and Wu, Dong and Yuan, Noah FQ and Nie, Zhengwei and Li, Jiarui and Sarker, Suchismita and Meng, Sheng and others},
  journal={Nat. Mater.},
  volume = {24},
  pages={1--7},
  year={2025},
  url= {https://doi.org/10.1038/s41563-025-02360-1}
}

@article{Xiao24,
 author = {Xiao, Kebin and Dong, Wen-Han and Wang, Xintong and Yu, Jiawei and Fu, Daran and Hu, Zhiqiang and Guo, Yunkai and Zhang, Qinqin and Hou, Xiaofei and Guo, Yanfeng and others},
 title = {Hidden Charge Order and Multiple Electronic Instabilities in EuTe4},
 journal = {Nano Lett.},
 volume = {24},
 number = {25},
 pages = {7681-7687},
 year = {2024},
 url = {https://doi.org/10.1021/acs.nanolett.4c01588}
}

@article{Meng22,
  title = {Angle-resolved photoemission spectroscopy study of charge density wave order in the layered semiconductor ${\text{EuTe}}_{4}$},
  author = {Zhang, Chen and Wu, Qi-Yi and Yuan, Ya-Hua and Zhang, Xin and Liu, Hao and Liu, Zi-Teng and Zhang, Hong-Yi and Song, Jiao-Jiao and Zhao, Yin-Zou and Wu, Fan-Ying and others},
  journal = {Phys. Rev. B},
  volume = {106},
  issue = {20},
  pages = {L201108},
  numpages = {5},
  year = {2022},
  month = {Nov},
  publisher = {American Physical Society},
  url = {https://link.aps.org/doi/10.1103/PhysRevB.106.L201108}
}

@article{Yang23,
  title = {Thermal hysteretic behavior and negative magnetoresistance in the charge density wave material ${\mathrm{EuTe}}_{4}$},
  author = {Zhang, Q. Q. and Shi, Y. and Zhai, K. Y. and Zhao, W. X. and Du, X. and Zhou, J. S. and Gu, X. and Xu, R. Z. and Li, Y. D. and Guo, Y. F. and others},
  journal = {Phys. Rev. B},
  volume = {107},
  issue = {11},
  pages = {115141},
  numpages = {7},
  year = {2023},
  month = {Mar},
  publisher = {American Physical Society},
  url = {https://link.aps.org/doi/10.1103/PhysRevB.107.115141}
}

@article{Bansal23,
  title = {Evolution of static charge density wave order, amplitude mode dynamics, and suppression of Kohn anomalies at the hysteretic transition in ${\mathrm{EuTe}}_{4}$},
  author = {Rathore, Ranjana and Pathak, Abhishek and Gupta, Mayanak K. and Mittal, Ranjan and Kulkarni, Ruta and Thamizhavel, A. and Singhal, Himanshu and Said, Ayman H. and Bansal, Dipanshu},
  journal = {Phys. Rev. B},
  volume = {107},
  issue = {2},
  pages = {024101},
  numpages = {9},
  year = {2023},
  month = {Jan},
  publisher = {American Physical Society},
  url = {https://link.aps.org/doi/10.1103/PhysRevB.107.024101}
}

@article{ThermoE25,
  title = {Enhanced Thermopower in a Magnetic Semiconductor ${\mathrm{Eu}\mathrm{Te}}_{4}$ with Multiple Charge-Density-Wave Instabilities},
  author = {Takahashi, Hidefumi and Yoshida, Kiichiro and Nakano, Akitoshi and Ishiwata, Shintaro},
  journal = {PRX Energy},
  volume = {4},
  issue = {3},
  pages = {033009},
  numpages = {9},
  year = {2025},
  month = {Aug},
  publisher = {American Physical Society},
  doi = {10.1103/9bvh-3r6w},
  url = {https://link.aps.org/doi/10.1103/9bvh-3r6w}
}

@article{Nonlocal23,
  title={Nonlocal Probing of Amplitude Mode Dynamics in Charge-Density-Wave Phase of EuTe4},
  author={Rathore, Ranjana and Singhal, Himanshu and Dwij, Vivek and Gupta, Mayanak K and Pathak, Abhishek and Chakera, Juzer Ali and Mittal, Ranjan and Roy, Aditya Prasad and Babu, Arun and Kulkarni, Ruta and others},
  journal={Ultrafast Sci.},
  volume={3},
  pages={0041},
  year={2023},
  url = {https://spj.science.org/doi/10.34133/ultrafastscience.0041}
}

@article{Oh25,
  title={Dynamics of a jointly commensurate moir{\'e} charge density wave},
  author={Oh, Kyoung Hun and Su, Yifan and Ning, Honglie and Lv, BQ and Zong, Alfred and Wu, Dong and Liu, Qiaomei and Kang, Gyeongbo and Choi, Hyeongi and Kim, Hyun-Woo J and others},
  journal={Nat. Commun.},
  volume={17},
  number={1},
  pages={10893},
  year={2026},
  url={https://www.nature.com/articles/s41467-026-77670-2}
}

@article{Ning25,
  title = {Bidirectional Ultrafast Control of Charge Density Waves via Phase Competition},
  author = {Ning, Honglie and Oh, Kyoung Hun and Su, Yifan and Shi, Zhengyan Darius and Wu, Dong and Liu, Qiaomei and Lv, B. Q. and Zong, Alfred and Kang, Gyeongbo and Choi, Hyeongi and others},
  journal = {Phys. Rev. Lett.},
  volume = {135},
  issue = {24},
  pages = {246504},
  numpages = {7},
  year = {2025},
  month = {Dec},
  publisher = {American Physical Society},
  doi = {10.1103/b1vl-qlkk},
  url = {https://link.aps.org/doi/10.1103/b1vl-qlkk}
}

@article{EC23,
  title={High-entropy ferroelectric materials},
  author={Qi, He and Chen, Liang and Deng, Shiqing and Chen, Jun},
  journal={Nat. Rev. Mater.},
  volume={8},
  number={6},
  pages={355--356},
  year={2023},
  url={https://www.nature.com/articles/s41578-023-00544-2}
}

@article{CALC22,
  title = {Orbital- and atom-dependent linear dispersion across the Fermi level induces charge density wave instability in ${\mathrm{EuTe}}_{4}$},
  author = {Pathak, Abhishek and Gupta, Mayanak K. and Mittal, Ranjan and Bansal, Dipanshu},
  journal = {Phys. Rev. B},
  volume = {105},
  issue = {3},
  pages = {035120},
  numpages = {8},
  year = {2022},
  month = {Jan},
  publisher = {American Physical Society},
  url = {https://link.aps.org/doi/10.1103/PhysRevB.105.035120}
}

@article{Nesting,
  title = {Fermi surface nesting and the origin of charge density waves in metals},
  author = {Johannes, M. D. and Mazin, I. I.},
  journal = {Phys. Rev. B},
  volume = {77},
  issue = {16},
  pages = {165135},
  numpages = {8},
  year = {2008},
  month = {Apr},
  publisher = {American Physical Society},
  url = {https://link.aps.org/doi/10.1103/PhysRevB.77.165135}
}

@article{FE1,
  title = {Ferroelectricity and Phase Transitions in Monolayer Group-IV Monochalcogenides},
  author = {Fei, Ruixiang and Kang, Wei and Yang, Li},
  journal = {Phys. Rev. Lett.},
  volume = {117},
  issue = {9},
  pages = {097601},
  numpages = {6},
  year = {2016},
  month = {Aug},
  publisher = {American Physical Society},
  doi = {10.1103/PhysRevLett.117.097601},
  url = {https://link.aps.org/doi/10.1103/PhysRevLett.117.097601}
}

@article{GL2001,
  title={Phase transitions in ferroelectrics: some historical remarks},
  author={Ginzburg, Vitalii L},
  journal={Phys.-Uspekhi},
  volume={44},
  number={10},
  pages={1037},
  year={2001},
  url={https://iopscience.iop.org/article/10.1070/PU2001v044n10ABEH001021}
}

@article{SI,
   journal={See Supplemental Material for details about: I. Computational methods; II. Origin of layer-selective polar CDW; III. Distinctions between ${\rm{EuTe}}_{4}$ and other $R{\rm{Te}}_{n}$; IV. Transition barriers and DFT basis for 1D effective model; V. Justifications for 1D effective model; VI. Negligible influence of vibrational entropy; VII. MC simulations for CDW states;  VIII. Justification for giant thermal hysteresis mechanism; and IX. Experimental predictions for validating our proposal, which includes Refs. \cite{VASP,PBE,CINEB,vdWD3,Born,TB06,TB08,COHP,COHP2,MLWF,W90,RTe2,RTe3,R2Te5,APL26,LV26,Spaldin,Phonopy1,Phonopy2,Pre1,Pre2,Pre3,Rozhkov25}}
}

@article{Entropy,
  title={Entropy-driven phase transitions},
  author={Frenkel, Daan},
  journal={Physica A},
  volume={263},
  number={1-4},
  pages={26--38},
  year={1999},
  url={https://www.sciencedirect.com/science/article/pii/S0378437198005019}
}

@article{Impurity,
  title = {Thermal hysteresis in the charge-density-wave transition of ${\mathrm{Lu}}_{5}{\mathrm{Rh}}_{4}{\mathrm{Si}}_{10}$},
  author = {Lue, C. S. and Kuo, Y.-K. and Hsu, F. H. and Li, H. H. and Yang, H. D. and Fodor, P. S. and Wenger, L. E.},
  journal = {Phys. Rev. B},
  volume = {66},
  issue = {3},
  pages = {033101},
  numpages = {4},
  year = {2002},
  month = {Jul},
  publisher = {American Physical Society},
  url = {https://link.aps.org/doi/10.1103/PhysRevB.66.033101}
}

@article{TaS219,
  title = {Origin of the Insulating Phase and First-Order Metal-Insulator Transition in $1T\text{\ensuremath{-}}{\mathrm{TaS}}_{2}$},
  author = {Lee, Sung-Hoon and Goh, Jung Suk and Cho, Doohee},
  journal = {Phys. Rev. Lett.},
  volume = {122},
  issue = {10},
  pages = {106404},
  numpages = {6},
  year = {2019},
  month = {Mar},
  publisher = {American Physical Society},
  url = {https://link.aps.org/doi/10.1103/PhysRevLett.122.106404}
}

@article{MTI,
  title = {Metastable metallic state and hysteresis below the metal-insulator transition in ${\mathrm{PrNiO}}_{3}$},
  author = {Granados, X. and Fontcuberta, J. and Obradors, X. and Torrance, J. B.},
  journal = {Phys. Rev. B},
  volume = {46},
  issue = {24},
  pages = {15683--15688},
  numpages = {0},
  year = {1992},
  month = {Dec},
  publisher = {American Physical Society},
  url = {https://link.aps.org/doi/10.1103/PhysRevB.46.15683}
}

@article{MultiQ1,
  title = {Atomistic Picture of Charge Density Wave Formation at Surfaces},
  author = {Wall, Simone and Krenzer, Boris and Wippermann, Stefan and Sanna, Simone and Klasing, Friedrich and Hanisch-Blicharski, Anja and Kammler, Martin and Schmidt, Wolf Gero and Horn-von Hoegen, Michael},
  journal = {Phys. Rev. Lett.},
  volume = {109},
  issue = {18},
  pages = {186101},
  numpages = {5},
  year = {2012},
  month = {Nov},
  publisher = {American Physical Society},
  url = {https://link.aps.org/doi/10.1103/PhysRevLett.109.186101}
}

@article{MultiQ2,
  title = {Hysteretic Melting Transition of a Soliton Lattice in a Commensurate Charge Modulation},
  author = {Hsu, Pin-Jui and Mauerer, Tobias and Vogt, Matthias and Yang, J. J. and Oh, Yoon Seok  and Cheong, S.-W. and Bode, Matthias and Wu, Weida},
  journal = {Phys. Rev. Lett.},
  volume = {111},
  issue = {26},
  pages = {266401},
  numpages = {5},
  year = {2013},
  month = {Dec},
  publisher = {American Physical Society},
  url = {https://link.aps.org/doi/10.1103/PhysRevLett.111.266401}
}

@article{Memristor,
  title={The growing memristor industry},
  author={Lanza, Mario and Pazos, Sebastian and Aguirre, Fernando and Sebastian, Abu and Le Gallo, Manuel and Alam, Syed M and Ikegawa, Sumio and Yang, J Joshua and Vianello, Elisa and Chang, Meng-Fan and others},
  journal={Nature},
  volume={640},
  number={8059},
  pages={613--622},
  year={2025},
  url={https://www.nature.com/articles/s41586-025-08733-5}
}

@article{EC25,
  title={Giant electrocaloric effect in high-polar-entropy perovskite oxides},
  author={Du, Feihong and Yang, Tiannan and Hao, Hua and Li, Shangshu and Xu, Chenhang and Yao, Tian and Song, Zhiwu and Shen, Jiahe and Bai, Chenyun and Luo, Ruhong and others},
  journal={Nature},
  volume={640},
  pages={924--930},
  year={2025},
  url={https://www.nature.com/articles/s41586-025-08768-8}
}

@article{EC21,
  title={High-entropy polymer produces a giant electrocaloric effect at low fields},
  author={Qian, Xiaoshi and Han, Donglin and Zheng, Lirong and Chen, Jie and Tyagi, Madhusudan and Li, Qiang and Du, Feihong and Zheng, Shanyu and Huang, Xingyi and Zhang, Shihai and others},
  journal={Nature},
  volume={600},
  number={7890},
  pages={664--669},
  year={2021},
  url={https://www.nature.com/articles/s41586-021-04189-5}
}

@article{VASP,
  title = {Efficient iterative schemes for ab initio total-energy calculations using a plane-wave basis set},
  author = {Kresse, G. and Furthm\"uller, J.},
  journal = {Phys. Rev. B},
  volume = {54},
  issue = {16},
  pages = {11169--11186},
  numpages = {0},
  year = {1996},
  month = {Oct},
  publisher = {American Physical Society},
  url = {https://link.aps.org/doi/10.1103/PhysRevB.54.11169}
}

@article{PBE,
  title = {Generalized Gradient Approximation Made Simple},
  author = {Perdew, John P. and Burke, Kieron and Ernzerhof, Matthias},
  journal = {Phys. Rev. Lett.},
  volume = {77},
  issue = {18},
  pages = {3865--3868},
  numpages = {0},
  year = {1996},
  month = {Oct},
  publisher = {American Physical Society},
  url = {https://link.aps.org/doi/10.1103/PhysRevLett.77.3865}
}

@article{CINEB,
    author = {Henkelman, Graeme and Uberuaga, Blas P. and Jónsson, Hannes},
    title = {A climbing image nudged elastic band method for finding saddle points and minimum energy paths},
    journal = {J. Chem. Phys.},
    volume = {113},
    number = {22},
    pages = {9901-9904},
    year = {2000},
    url = {https://doi.org/10.1063/1.1329672}
}

@article{vdWD3,
    author = {Grimme, Stefan and Antony, Jens and Ehrlich, Stephan and Krieg, Helge},
    title = {A consistent and accurate ab initio parametrization of density functional dispersion correction (DFT-D) for the 94 elements H-Pu},
    journal = {J. Chem. Phys.},
    volume = {132},
    number = {15},
    pages = {154104},
    year = {2010},
    month = {04},
    issn = {0021-9606},
    url = {https://doi.org/10.1063/1.3382344}
}

@article{Born,
  title = {First-principles study of spontaneous polarization in multiferroic $\mathrm{Bi}\mathrm{Fe}{\mathrm{O}}_{3}$},
  author = {Neaton, J. B. and Ederer, C. and Waghmare, U. V. and Spaldin, N. A. and Rabe, K. M.},
  journal = {Phys. Rev. B},
  volume = {71},
  issue = {1},
  pages = {014113},
  numpages = {8},
  year = {2005},
  month = {Jan},
  publisher = {American Physical Society},
  url = {https://link.aps.org/doi/10.1103/PhysRevB.71.014113}
}

@article{TB06,
  title = {Theory of stripes in quasi-two-dimensional rare-earth tellurides},
  author = {Yao, Hong and Robertson, John A. and Kim, Eun-Ah and Kivelson, Steven A.},
  journal = {Phys. Rev. B},
  volume = {74},
  issue = {24},
  pages = {245126},
  numpages = {7},
  year = {2006},
  month = {Dec},
  publisher = {American Physical Society},
  url = {https://link.aps.org/doi/10.1103/PhysRevB.74.245126}
}

@article{TB08,
  title = {Angle-resolved photoemission study of the evolution of band structure and charge density wave properties in $R{\text{Te}}_{3}$ ($R=\text{Y}$, La, Ce, Sm, Gd, Tb, and Dy)},
  author = {Brouet, V. and Yang, W. L. and Zhou, X. J. and Hussain, Z. and Moore, R. G. and He, R. and Lu, D. H. and Shen, Z. X. and Laverock, J. and Dugdale, S. B. and others},
  journal = {Phys. Rev. B},
  volume = {77},
  issue = {23},
  pages = {235104},
  numpages = {16},
  year = {2008},
  month = {Jun},
  publisher = {American Physical Society},
  url = {https://link.aps.org/doi/10.1103/PhysRevB.77.235104}
}

@article{COHP,
  title={Crystal orbital Hamilton populations (COHP): energy-resolved visualization of chemical bonding in solids based on density-functional calculations},
  author={Dronskowski, Richard and Bloechl, Peter E},
  journal={J. Phys. Chem.},
  volume={97},
  number={33},
  pages={8617--8624},
  year={1993},
  publisher={ACS Publications},
  url= {https://pubs.acs.org/doi/abs/10.1021/j100135a014}
}

@article{COHP2,
  title={Crystal orbital Hamilton population (COHP) analysis as projected from plane-wave basis sets},
  author={Deringer, Volker L and Tchougr{\'e}eff, Andrei L and Dronskowski, Richard},
  journal={J. Phys. Chem. A},
  volume={115},
  number={21},
  pages={5461--5466},
  year={2011},
  publisher={ACS Publications},
  url= {https://pubs.acs.org/doi/abs/10.1021/jp202489s}
}

@article{MLWF,
  title = {Maximally localized generalized Wannier functions for composite energy bands},
  author = {Marzari, Nicola and Vanderbilt, David},
  journal = {Phys. Rev. B},
  volume = {56},
  issue = {20},
  pages = {12847--12865},
  numpages = {0},
  year = {1997},
  month = {Nov},
  publisher = {American Physical Society},
  doi = {10.1103/PhysRevB.56.12847},
  url = {https://link.aps.org/doi/10.1103/PhysRevB.56.12847}
}

@article{W90,
  title={wannier90: A tool for obtaining maximally-localised Wannier functions},
  author={Mostofi, Arash A and Yates, Jonathan R and Lee, Young-Su and Souza, Ivo and Vanderbilt, David and Marzari, Nicola},
  journal={Comput. Phys. Commun.},
  volume={178},
  number={9},
  pages={685--699},
  year={2008},
  publisher={Elsevier},
  url = {https://www.sciencedirect.com/science/article/abs/pii/S0010465507004936}
}

@article{RTe2,
  title = {Electronic structure and charge-density wave formation in $\mathrm{La}{\mathrm{Te}}_{1.95}$ and $\mathrm{Ce}{\mathrm{Te}}_{2.00}$},
  author = {Shin, K. Y. and Brouet, V. and Ru, N. and Shen, Z. X. and Fisher, I. R.},
  journal = {Phys. Rev. B},
  volume = {72},
  issue = {8},
  pages = {085132},
  numpages = {9},
  year = {2005},
  month = {Aug},
  publisher = {American Physical Society},
  url = {https://link.aps.org/doi/10.1103/PhysRevB.72.085132}
}

@article{RTe3,
author = {Yumigeta, Kentaro and Qin, Ying and Li, Han and Blei, Mark and Attarde, Yashika and Kopas, Cameron and Tongay, Sefaattin},
title = {Advances in Rare-Earth Tritelluride Quantum Materials: Structure, Properties, and Synthesis},
journal = {Adv. Sci.},
volume = {8},
number = {12},
pages = {2004762},
year = {2021},
url = {https://advanced.onlinelibrary.wiley.com/doi/abs/10.1002/advs.202004762}
}

@article{R2Te5,
  title = {Charge density wave formation in ${R}_{2}{\mathrm{Te}}_{5}$ ($R=\mathrm{Nd}$, Sm, and Gd)},
  author = {Shin, K. Y. and Laverock, J. and Wu, Y. Q. and Condron, C. L. and Toney, M. F. and Dugdale, S. B. and Kramer, M. J. and Fisher, I. R.},
  journal = {Phys. Rev. B},
  volume = {77},
  issue = {16},
  pages = {165101},
  numpages = {9},
  year = {2008},
  month = {Apr},
  publisher = {American Physical Society},
  url = {https://link.aps.org/doi/10.1103/PhysRevB.77.165101}
}

@article{APL26,
  title={Termination-dependent electronic structure reconstruction in epitaxial EuTe4 thin films},
  author={Zhou, Yaling and Yu, Fan and Zhong, Jianqi and Zhou, Jinming and Wang, Gan and Xu, Gang and Zhang, Yi},
  journal={Appl. Phys. Lett.},
  volume={128},
  number={23},
  year={2026},
  url={https://doi.org/10.1063/5.0333900}
}

@article{Spaldin,
   author = {Bhowal, Sayantika and Spaldin, Nicola A.},
   title = {Polar Metals: Principles and Prospects}, 
   journal= {Annu. Rev. Mater. Res.},
   year = {2023},
   volume = {53},
   pages = {53-79},
   url = {https://www.annualreviews.org/content/journals/10.1146/annurev-matsci-080921-105501}
  }
\end{document}